# Quantile based global sensitivity measures

S. Kucherenko*, S. Song

*Imperial College London, London, SW7 2AZ, UK*

*e-mail: s.kucherenko@imperial.ac.uk*

**Abstract.** New global sensitivity measures based on quantiles of the output are introduced. Such measures can be used for global sensitivity analysis of problems in which $\alpha$-th quantiles are explicitly the functions of interest and for identification of variables which are the most important in achieving extreme values of the model output. It is proven that there is a link between introduced measures and Sobol' main effect sensitivity indices. Two different Monte Carlo estimators are considered. It is shown that the double loop reordering approach is much more efficient than the brute force estimator. Several test cases and practical case studies related to structural safety are used to illustrate the developed method. Results of numerical calculations show the efficiency of the presented technique.

1. Introduction

Global sensitivity analysis (GSA) is the study of how the uncertainty in model output is apportioned to the uncertainty in model inputs. It enables the identification of key parameters whose uncertainty most affects the output. It can be used to rank variables, fix unessential variables and reduce model complexity. Over the years there has been a significant progress in developing global sensitivity measures which quantify the uncertainty of inputs in the uncertainty of outputs [1].

Variance-based method also known as the Sobol' method of global sensitivity indices is one of the most efficient and comprehensive GSA techniques [2]. However, generally variance-based methods require a large number of function evaluations to achieve acceptable convergence. They can become impractical for large engineering problems. Besides, being centered around the first and the second order moments of the output distribution function these methods are not well suited for GSA of problems in which higher moments or other statistical values can play a significant role. These methods also cannot identify variables which are the most important in achieving extreme values of the model output.

A number of alternative GSA techniques have been proposed recently. Derivative based global sensitivity measures (DGSM) have been introduced by Kucherenko and co-authors [3]. Sobol' and Kucherenko [4] proved theoretically that there is a link between DGSM and Sobol' total sensitivity indices. Variance based measures by definition are moment dependent. Borgonovo [5] proposed a moment independent measure.

There are problems in which analysts are interested only in certain regions of output values. Examples include problems of mechanical engineering (f.e beam under loading), civil engineering (f.e the reliability of buildings under seismic load), environmental science (f.e a dose of contamination of soil, water, air ), chemical engineering (f.e stability behavior of dynamic systems such as chemical reactors). Reliability analysis of system failure in such systems is often based on computing the probability of failure with respect to some performance function g(x), where x is a vector of uncertain variables. In this formulation, g(x) < C denotes the failure state, while g(x) > C denotes the safe state and g(x) = C is known as the limit state. It is possible to reformulate this problem in terms of critical $\alpha$ quantile of the cumulative distribution function (CDF) of the performance function. There are also problems in which $\alpha$-th quantiles of CDF of the output are used explicitly (reliability analysis, risk analysis in finance, etc ). For such problems conventional sensitivity measures are not adequate.

In this paper we introduce sensitivity measures based on quantiles of the output. Such measures can be used for global sensitivity analysis of problems in which $\alpha$-th quantiles are the functions of interest and for problems in which the analysts are interested in ranking of inputs contributing to the extreme values of the output. It is shown that there is a direct link between introduced measures and Sobol' main effect sensitivity indices. We consider two different MC estimators of the introduced sensitivity measures and present the results of numerical tests including two practical case studies related to structural safety.

We note, that accurate and numerically efficient computation of quantiles is a difficult mathematical problem. Importance sampling is a widely used technique for variance reduction of Monte Carlo estimates. The basic idea of importance sampling is to change the sampling distribution so that a greater concentration of samples is generated in a region of the sample space which has a dominant impact on the calculations. Importance sampling has been successfully applied in rare-event simulation. Glynn suggested to use importance sampling for computing extreme quantiles [6]. He proved a central limit theorem for proposed importance sampling quantile estimators and provided efficiency comparisons in a certain asymptotic setting. A sample-based quantile estimators with adaptive importance sampling was introduced in [7]. Oakley proposed an efficient technique for estimating percentiles of models with uncertain parameters related to water drain design [8]. Design is based on predicting extreme events that are related to the $95^{th}$ percentile of the cumulative distribution function of the model output.

This paper is organized as follows: The next Section presents a brief review of Sobol' sensitivity indices. We consider both deterministic and probabilistic approaches. In Section 3 two different sensitivity measures based on quantiles of output are introduced. The brute force MC estimator and the estimator based on the double loop reordering approach are discussed in Section 4. In Section 5 we present Value at Risk which is a measure widely used in financial risk analysis and show its link with importance measures based on quantiles of model output. Linear models with normally distributed variables are considered in Section 6. In this Section we prove a Theorem establishing a direct link between introduced measures and Sobol' main effect sensitivity indices. The results of numerical tests for models given analytically are considered in Section 7. Section 8 presents the results of practical case studies linked to reliability analysis. Finally, conclusions are summarized in Section 9.

## 2. Sobol' sensitivity indices

## 2.1. Deterministic approach

Consider an integrable function $f(\mathbf{x})$ defined in the unit hypercube $H^d$. It can be expanded in the following form:

$$f(\mathbf{x}) = f_0 + \sum_{s=1}^{d} \sum_{i_1 < \ldots < i_s} f_{i_1 \ldots i_s}(x_{i_1}, \ldots, x_{i_s}). \tag{1}$$

Expansion (1) is a sum of $2^d$ components. It can also be presented as

$$f(\mathbf{x}) = f_0 + \sum_{i} f_i(x_i) + \sum_{i<j} f_{ij}(x_i, x_j) + \ldots + f_{12\ldots d}(x_1, x_2, \ldots, x_d).$$

Each of the components $f_{i_1 \ldots i_s}(x_{i_1}, \ldots, x_{i_s})$ is a function of a unique subset of variables from $\mathbf{x}$. The components $f_i(x_i)$ are called first order terms, $f_{ij}(x_i, x_j)$ - second order terms and so on.

It was proven in [9] that the expansion (1) is unique if

$$\int_0^1 f_{i_1 \ldots i_s}(x_{i_1}, \ldots, x_{i_s}) dx_{i_k} = 0, \ 1 \le k \le s, \tag{2}$$

in which case it is called a decomposition into summands of different dimensions. This decomposition is also known as the ANOVA (ANalysis Of VAriances) decomposition. The ANOVA decomposition is orthogonal, *i.e.* for any two subsets $u \ne v$ an inner product

$$\int_0^1 f_u(x) f_v(x) dx = 0.$$

It follows from expansions (1) and (2) that

$$\int_0^1 f(x) dx_1 \ldots dx_n = f_0, \quad \int_0^1 f(x) \prod_{k \ne i} dx_k = f_0 + f_i(x_i),$$

$$\int_0^1 f(x) \prod_{k \ne i, j} dx_k = f_0 + f_i(x_i) + f_j(x_j) + f_{i,j}(x_i, x_j) \ldots \tag{3}$$

and so on.

For square integrable functions, the variances of the terms in the ANOVA decomposition add up to the total variance of the function

$$V = \sum_{s=1}^{d} \sum_{i_1 < \cdots < i_s} V_{i_1 \ldots i_s},$$

where $V_{i_1 \ldots i_s} = \int_0^1 f_{i_1 \ldots i_s}^2(x_{i_1}, \ldots, x_{i_s}) dx_{i_1}, \ldots, x_{i_s}$ are known as partial variances.

Sobol' defined global sensitivity indices as the ratios

$$S_{i_1...i_s} = V_{i_1...i_s} / V.$$

All $S_{i_1...i_s}$ are non negative and add up to one

$$\sum_{s=1}^{n} \sum_{i_1 < ... < i_s} S_{i_1...i_s} = 1.$$

$S_{i_1...i_s}$ can be viewed as a natural sensitivity measure of a set of variables $x_{i_1},...,x_{i_s}$. In the general case all global sensitivity indices can be important. Their straightforward calculation using the ANOVA decomposition would result in $2^d$ integral evaluations of the summands $f_{i_1...i_s}(x_{i_1},...,x_{i_s})$ using (3) and $2^d$ integral evaluations for calculations of $V_{i_1...i_s}$. For high dimensional problems such an approach is impractical. Sobol' introduced sensitivity indices for subsets of variables [9].

Consider an arbitrary subset of the variables $y = (x_{i_1},...,x_{i_s})$, $1 \leq i_1 < ... < i_s \leq d$, $K = (i_1,...,i_s)$, $1 \leq s < d$ and a complementary subset $z = (x_{i_1},...,x_{i_{d-s}})$, so that $x = (y, z)$. The variance corresponding to $y$ is defined as

$$V_y = \sum_{m=1}^{s} \sum_{(i_1 < \cdots < i_m) \in K} V_{i_1...i_m}.$$

$V_y$ includes all partial variances $V_{i_1}$, $V_{i_2}$,..., $V_{i_1...i_m}$ such that their subsets of indices $(i_1,...,i_m) \in K$. Homma and Saltelli [10] introduced the total variance $V_y^{tot}$:

$$V_y^{tot} = V - V_z.$$

$V_y^{tot}$ consists of all $V_{i_1...i_m}$ such that at least one index $i_p \in K$ while the remaining indices can belong to the complimentary to $K$ set $\bar{K}$. The corresponding global sensitivity indices are defined as

$$\begin{aligned} S_y &= V_y / V, \\ S_y^{tot} &= V_y^{tot} / V. \end{aligned} \quad (4)$$

Total sensitivity indices $S_y^{tot}$ are used to identify non-important variables which can then be fixed at their nominal values to reduce model complexity.

Collectively $S_y$, $S_y^{tot}$ in the case of independent variables are known as Sobol' sensitivity indices. One of the most important results obtained by Sobol' is an effective way of computing sensitivity indices using direct formulas in a form of high-dimensional integrals [11]. Sobol' formulas were further improved in a number of papers ( see f.e. [12, 13]).

The important indices in practice are $S_i$ and $S_i^{tot}$. Their knowledge in most cases provides sufficient information to determine the sensitivity of the analyzed function to individual input variables. The use of $S_i$ and $S_i^{tot}$ reduces the number of index calculations from $O(2^d)$ to just $O(2d)$.

### 2.2. Probabilistic approach

Consider a model function $Y = f(x_1,...,x_d)$ defined in $R^d$ with an input vector $x = (x_1,...,x_d)$. Here is $x$ a real-valued random variable with a continuous probability distribution function (PDF) $\rho(x_1,...,x_d)$. It is assumed that $f(x_1,...,x_d)$ has a finite variance. Consider an arbitrary subset of the variables $y = (x_{i_1},...,x_{i_s})$, $1 \leq s < d$ and a complementary subset $z = (x_{i_1},...,x_{i_{d-s}})$, so that x=(y,z). Consider the first order $S_i$ sensitivity index for input $x_i$. In this case $y = (x_i)$, $1 \leq i \leq d$. A commonly used notation for a complimentary set $z$ is $z = x_{\sim i}$.

The total variance of $f(x_1,...,x_d)$ can be decomposed as

$$V = V_{X_i}\left(E_{X_{\sim i}}(Y|X_i)\right) + E_{X_i}\left(V_{X_{\sim i}}(Y|X_i)\right). \tag{5}$$

Here $X_i$ is a realization of the random variable $x_i$, $E_{X_{\sim i}}(Y|X_i)$ and $V_{X_{\sim i}}(Y|X_i)$ are a conditional expectation and a conditional variance of the output $Y$ with $x_i = X_i$, respectively. Normalized by the total variance, this expression leads to the equality

$$1 = \frac{V_{X_i}\left(E_{X_{\sim i}}(Y|X_i)\right)}{V} + \frac{E_{X_i}\left(V_{X_{\sim i}}(Y|X_i)\right)}{V}.$$

The first order $S_i$ sensitivity index for one input has a form:

$$S_i = \frac{V_{X_i}\left(E_{X_{\sim i}}(Y|X_i)\right)}{V}, \tag{6}$$

while $S_i^{tot}$ is defined as

$$S_i^{tot} = \frac{E_{X_{\sim i}}\left(V_{X_i}(Y|X_{\sim i})\right)}{V}. \tag{7}$$

Here $S_i$ is defined in terms of variances of conditional expectation and $S_i^{tot}$ is defined in terms of expectation of conditional variance are the same sensitivity indices as defined by (4). We note, that the importance measure (6) (although not normalized by $V$) was firstly suggested by Iman and Hora [14].

### 3. Sensitivity measures based on quantiles of output

Quantiles of the output CDF are used in reliability analysis, risk analysis in finance and some other areas. For such problems conversional sensitivity measures are not adequate.

The $\alpha$-th quantile of the output CDF $q_Y(\alpha)$ by definition is

$$\alpha = \int_{-\infty}^{q_Y(\alpha)} \rho_Y(y)dy = P\{Y \leq q_Y(\alpha)\} \tag{8}$$

or more formally

$$q_Y(\alpha) = F_Y^{-1}(\alpha) = \inf\{y | F(Y \leq y) \geq \alpha\}, \tag{9}$$

where $\rho_Y(y)$ and $F_Y(y)$ are PDF and CDF of the output $Y$, respectively. Equation (8) can be presented in a different form which is more useful for practical use:

$$\alpha = \int_{Y \leq q_Y(\alpha)} \rho_X(x)dx = \int_{\Omega} I_{Y \leq q_Y(\alpha)}[Y(x)] \rho_X(x)dx,$$

where $I(x)$ is an indicator function.

The moment independent measure based on quantiles was proposed by Chun *et al* in [15]:

$$CHT_i = \left[\int (q_{Y|X_i}(\alpha) - q_Y(\alpha))^2 d\alpha\right]^{1/2} \bigg/ E[Y]. \tag{10}$$

Here $q_{Y|X_i}(\alpha)$ is $\alpha$-th quantile of CDF of the output conditional on the input variable $x_i$ being fixed at $x_i = X_i$. Measure $CHT_i$ does not depend on $\alpha$ and it can be formally written as

$$CHT_i = (E_\alpha[(q_{Y|X_i}(\alpha) - q_Y(\alpha))^2])^{1/2} / E[Y].$$

We define new quantile-based sensitivity measures $\bar{q}_i^{(1)}(\alpha)$ and $\bar{q}_i^{(2)}(\alpha)$:

$$\bar{q}_i^{(1)}(\alpha) = E_{x_i}\left(|q_Y(\alpha) - q_{Y|X_i}(\alpha)|\right), \tag{11}$$

$$\bar{q}_i^{(2)}(\alpha) = E_{x_i}\left[\left(q_Y(\alpha) - q_{Y|X_i}(\alpha)\right)^2\right]. \tag{12}$$

We note that measures $\bar{q}_i^{(1)}(\alpha)$ and $\bar{q}_i^{(2)}(\alpha)$ can be seen as an expectation over the range of $x_i$ of a metric ($L^1$ norm or $C_1$ distance) and square of the metric ($L^2$ norm or $L^2$ distance) on a set of $q_Y(\alpha)$ and $q_{Y|X_i}(\alpha)$, correspondingly.

We also introduce normalized versions of quantile-based sensitivity measures $Q_i^{(1)}(\alpha)$ and $Q_i^{(2)}(\alpha)$:

$$Q_i^{(1)}(\alpha) = \frac{\overline{q}_i^{(1)}(\alpha)}{\sum_{j=1}^{d} \overline{q}_j^{(1)}(\alpha)}, \qquad (13)$$

$$Q_i^{(2)} = \frac{\overline{q}_i^{(2)}(\alpha)}{\sum_{j=1}^{d} \overline{q}_j^{(2)}(\alpha)}. \qquad (14)$$

We notice that $\{Q_i^{(1)}(\alpha), Q_i^{(2)}(\alpha)\} \in [0,1]$ by construction. These measures can be evaluated explicitly as

$$\overline{q}_i^{(1)}(\alpha) = E_{x_i}\left(|q_Y(\alpha) - q_{Y|X_i}(\alpha)|\right) = \int |q_Y(\alpha) - q_{Y|X_i}(\alpha)|\, dF_{x_i}, \qquad (15)$$

$$\overline{q}_i^{(2)}(\alpha) = E_{x_i}\left[\left(q_Y(\alpha) - q_{Y|X_i}(\alpha)\right)^2\right] = \int \left(q_Y(\alpha) - q_{Y|X_i}(\alpha)\right)^2 dF_{x_i}. \qquad (16)$$

We note a formal structural similarity in definitions of $\overline{q}_i^{(1)}(\alpha)$ and $\overline{q}_i^{(2)}(\alpha)$ and a moment independent measure proposed by Borgonovo [5]:

$$\delta_i = \frac{1}{2} E_{x_i}[s(x_i)] = \frac{1}{2} \int \left[\int |\rho_Y(y) - \rho_{Y|X_i}(y)|\, dy\right] dF_{x_i}, \qquad (17)$$

where $\rho_{Y|X_i}(y)$ is the conditional on $x_i = X_i$ PDF of the output $Y$.

## 4. Monte Carlo estimators

### 4.1. The brute force estimator

CDF $F_Y(y)$ of the output $Y$ is rarely known explicitly. In practice it is computed numerically by sampling $N$ point $\boldsymbol{x}^l = (x_1^l, ..., x_d^l)$, $l = 1, ..., N$ using MC or QMC sampling methods and estimating CDF as

$$F_Y^{(N)}(y) = \frac{1}{N} \sum_{l=1}^{N} I(Y(x_1^l, ..., x_d^l) < y). \qquad (18)$$

Similarly CDF $F_{Y|X_i}(y)$ of the output conditional on the input variable $x_i$ being fixed at $x_i = X_i$ is estimated as

$$F_{Y|X_i}^{(N)}(y) = \frac{1}{N}\sum_{l=1}^{N} I(Y(x_1^l,...,x_i = X_i,...,x_d^l) < y) \qquad (19)$$

Once CDF's are computed, MC/QMC estimates of quantiles can be computed as

$$q_Y^{(N)}(\alpha) = [F_Y^{(N)}]^{-1}(\alpha) = \inf\{y | F_Y^{(N)}(Y \le y) \ge \alpha\}, \qquad (20)$$

$$q_{Y|X_i}^{(N)}(\alpha) = [F_{Y|X_i}^{(N)}]^{-1}(\alpha) = \inf\{y | F_{Y|X_i}^{(N)}(Y \le y | x_i = X_i) \ge \alpha\}. \qquad (21)$$

The Monte Carlo (MC) or Quasi-Monte Carlo (QMC) estimators $\widehat{\bar{q}}_i^{(1)}(\alpha)$, $\widehat{\bar{q}}_i^{(2)}(\alpha)$ given by (15) and (16) have the form:

$$\widehat{\bar{q}}_i^{(1)}(\alpha) = \frac{1}{N}\sum_{j=1}^{N} |q_Y^{(N)}(\alpha) - q_{Y|x_i=X_i^{'(j)}}^{(N)}(\alpha)|, \qquad (22)$$

$$\widehat{\bar{q}}_i^{(2)}(\alpha) = \frac{1}{N}\sum_{j=1}^{N} \left(q_Y^{(N)}(\alpha) - q_{Y|x_i=X_i^{'(j)}}^{(N)}(\alpha)\right)^2. \qquad (23)$$

In practice for the $j$-th trial we generate two independent points $\boldsymbol{x}^j = (x_1^j,...,x_d^j)$ and $\boldsymbol{x}^{'(j)} = (x_1^{'(j)},...,x_d^{'(j)})$, $j = 1,...,N$. We use set $\{\boldsymbol{x}^j\}$ distributed according to a joint PDF $\rho(x_1,...,x_d)$ for estimating $q_Y^{(N)}(\alpha)$ using (18), (20) and a mixed $\{(x_{\sim i}^j, x_i^{'(j)})\}$ set distributed according to a conditional PDF $\rho(x_1,...,x_d | x_i = X_i^{'(j)})$, where $(x_{\sim i}^j, x_i^{'(j)}) = (x_1^j,...,x_i^{'(j)},...,x_d^j)$ for estimating $q_{Y|x_i=X_i^{'(j)}}(\alpha)$ using (19), (21).

The total number of sampled points and function evaluations for a fixed $i$ is equal to $N_T^i = NN + N$. To compute all $d$ estimates $N_T = dNN + N = N(dN+1)$. To achieve a good convergence $N$ should be large, which means that $N_T$ depends on $N$ quadratically, hence such a simple brute force algorithm requires high computational efforts.

Oakley proposed an efficient technique for estimating percentiles of models with uncertain parameters in which the output as a function of its inputs is modelled as a Gaussian process [8]. A few initial runs of the model are used for building a metamodel, to choose further suitable design points and to make inferences about the percentile of interest.

### 4.2. Double loop reordering (DLR) approach

In this section we consider a different and more efficient estimator: $N$ points $x^{(j)}$, $j=1,2,...,N$ are generated from the joint PDF and values of the output $\{Y^{(j)}\}$ are found at this sampled set $\{x^{(j)}\}$.

CDF $F_Y(y)$ of the output Y is found as before using (18). For each random variable $x_i$, the sample set $x^{(j)}$, $j=1,2,...,N$ with corresponding values of $\{Y^{(j)}\}$ is sorted in ascending order with respect to the values of $x_i$ and subdivided in $M$ equally populated partitions (bins) with $N_m = N/M$ points in each bin ($M < N$). Within each bin the CDF $F_{Y|X_i}(y)$ of the output conditional on the input variable $x_i$ being fixed at $x_i = X_i^l$, $l=1,...,N_m$ is estimated as

$$F_{Y|X_i}^{(N_m)}(y) = \frac{1}{N_m} \sum_{l=1}^{N_m} I(Y(x_1^l,...,x_i^l = X_i^l,...,x_d^l) < y). \tag{24}$$

Once CDF's are computed, MC/QMC estimates of quantiles can be computed as before using (20), (21). Finally, the MC/QMC estimators of integrals (15), (16) have the form:

$$\widehat{\overline{q}}_i^{(1)}(\alpha) = \frac{1}{M} \sum_{j=1}^{M} |q_Y^{(N_m)}(\alpha) - q_{Y|x_i=X_i^{(j)}}^{(N_m)}(\alpha)|, \tag{25}$$

$$\widehat{\overline{q}}_i^{(2)}(\alpha) = \frac{1}{M} \sum_{j=1}^{M} \left(q_Y^{(N_m)}(\alpha) - q_{Y|x_i=X_i^{(j)}}^{(N_m)}(\alpha)\right)^2. \tag{26}$$

The subdivision in bins is done in the same way for all inputs using the same set of sampled points. Further we call this method the double loop reordering (DLR) approach. The total number of sampled points and function evaluations required for computing $\widehat{\overline{q}}_i^{(1)}(\alpha)$ and $\widehat{\overline{q}}_i^{(2)}(\alpha)$ for all $i=1,2,...,d$ is $N_T = N$. Apparently this is a much more efficient estimator than the brute force estimator. It dramatically reduces the number of samples required to achieve a given level of accuracy. A similar estimator was proposed in [16, 17].

### 5. Value at Risk

Finance is one of the potential areas in which the proposed new sensitivity measures can be used. In this Section we briefly consider a risk measure widely used in risk analysis in finance, namely Value at Risk (VaR). VaR is the amount of potential loss with given probability over the specific time period. VaR is a technique used to measure and quantify the level of financial risk within a firm or investment portfolio.

Consider tomorrow's price $P$ for a portfolio which depends on a set of $d$ random variables, specifically, tomorrow's values $\{x_i\}$, $i=1,...,d$ for a set of "risk factors" (those risk factors might be interest rates, commodity prices, equity prices, etc.). The portfolio contains $m$ contracts whose values tomorrow are $\{C_k\}$, $k=1,...,m$. Contract values are a function $C_k(x_1,...,x_d)$ of the risk factors, hence the portfolio price function has a form:

$$P(x_1, x_2, ..., x_d) = \sum_{k=1}^{m} C_k(x_1, x_2, ..., x_d).$$

We are interested in the value of the $\alpha$-quantile of the profit and loss (P&L) distribution function of the portfolio which we denote $f_Y(y)$, where $Y = \Delta P$ is a change in the portfolio value over a specific defined time period $T$. VaR is an $\alpha$-quantile of model output as defined by equation (9), where the model output is the P&L of a portfolio.

We can construct the quantile measures based on VaR for risk management problems, which we denote $Q_i^{VaR}$:

$$Q_i^{VaR} = \frac{E_{X_i}\left(d[VaR_Y(\alpha) - VaR_{Y|X_i}(\alpha)]\right)}{\sum_{j=1}^{d} E_{X_i}\left(d[VaR_Y(\alpha) - VaR_{Y|X_j}(\alpha)]\right)}.$$

As before we will use two types of distances $d[]$: $C_1$ distance and $L_2$ distance. We notice that $Q_i^{VaR}$ are defined similarly to sensitivity measures $Q_i$ (13), (14).

### 6. Linear model with normally distributed variables

Consider the following model

$$Y = a_1 x_1 + a_2 x_2 + ... + a_d x_d, \tag{27}$$

where $x_1, x_2, ..., x_d$ are independently distributed normal random variables, i.e. $x_i \sim N(\mu_i, \sigma_i^2)$ and $a_1, a_2, ..., a_d$ are constant coefficients such that not all of them are equal to 0. The PDF of the output $Y$ is a normal distribution $Y \sim N\left(\sum_{i=1}^{d} a_i \mu_i, \sum_{i=1}^{d} a_i^2 \sigma_i^2\right)$, and the conditional PDF is

$$Y | X_i \sim N\left(a_i x_i + \sum_{j=1, j\neq i}^{d} a_j \mu_j, \sum_{j=1, j\neq i}^{d} a_j^2 \sigma_j^2\right).$$ The values of Sobol' sensitivity indices are $S_i = S_i^{tot} = \frac{a_i^2 \sigma_i^2}{\sum_{j=1}^{d} a_j^2 \sigma_j^2}$, that is only the variance of $x_i$ defines the influence of $x_i$ on the output $Y$.

For this model

$$q_Y(\alpha) = \sum_{i=1}^{d} a_i \mu_i + \Phi^{-1}(\alpha) \sqrt{\sum_{i=1}^{d} a_i^2 \sigma_i^2},$$

$$q_{Y|X_i = x_i}(\alpha) = a_i x_i + \sum_{j=1, j\neq i}^{d} a_j \mu_j + \Phi^{-1}(\alpha) \sqrt{\sum_{j=1, j\neq i}^{d} a_j^2 \sigma_j^2},$$

where $\Phi^{-1}(\alpha)$ is the inverse error function. Hence

$$q_Y(\alpha) - q_{Y|X_i=x_i}(\alpha) = a_i(\mu_i - x_i) + \Phi^{-1}(\alpha)\left(\sqrt{\sum_{i=1}^{d} a_i^2 \sigma_i^2} - \sqrt{\sum_{j=1, j \neq i}^{d} a_j^2 \sigma_j^2}\right).$$

Applying formula (12) we obtain

$$q_i^{(2)}(\alpha) = E_{X_i}\left[\left(a_i(\mu_i - x_i) + \Phi^{-1}(\alpha)\left(\sqrt{\sum_{i=1}^{d} a_i^2 \sigma_i^2} - \sqrt{\sum_{j=1, j \neq i}^{d} a_j^2 \sigma_j^2}\right)\right)^2\right].$$

After transformation we arrive at

$$q_i^{(2)}(\alpha) = a_i^2 \sigma_i^2 + [\Phi^{-1}(\alpha)]^2 \left(\sqrt{\sum_{i=1}^{d} a_i^2 \sigma_i^2} - \sqrt{\sum_{j=1, j \neq i}^{d} a_j^2 \sigma_j^2}\right)^2. \quad (28)$$

**Theorem 1.** For the linear additive model (27) with normally distributed variables

$$Q_i^{(2)}(\alpha = 0.5) = S_i. \quad (29)$$

**Proof.** We note that $\Phi^{-1}(\alpha = 0.5) = 0$. Then $q_i^{(2)}(\alpha) = a_i^2 \sigma_i^2$ and (29) follows directly from formula (28) and definition of $Q_i^{(2)}$ (14).

## 7. Test models given analytically

In this Section we present several test cases given analytically to illustrate the developed method.

### 7.1. Independent inputs

For the first two cases considered in this Section we found analytical values of sensitivity indices, so that they can be used as benchmarks for verification of numerical estimates.

**Test 1.** Consider the following model

$$Y = a_1 x_1 + a_2 x_2 + \ldots + a_d x_d, \quad (30)$$

where $x_1, x_2, \ldots, x_d$ are independently distributed normal random variables, i.e. $x_i \sim N(\mu_i, \sigma_i^2)$ and $a_1, a_2, \ldots, a_d$ are constant coefficients such that not all of them are equal to 0. The PDF of the output $Y$, the conditional PDF and the values of Sobol' indices were presented in Section 6.

Further we consider the case when d=4 and the mean values and standard deviations are $\boldsymbol{\mu} = (1, 3, 5, 7)$ and $\boldsymbol{\sigma} = (1, 1.5, 2, 2.5)$ respectively. All coefficients $a_i$ =1. The values of sensitivity indices $S_i = S_i^{tot}$ ={0.0741, 0.167, 0.296, 0.463}. The values of Borgonovo's measure $\delta_i$ ={0.093, 0.1439, 0.202, 0.273}. All values are computed analytically.

Values of quantile measures $Q_i^{(1)}$ and $Q_i^{(2)}$ versus $\alpha$ and values of sensitivity indices are presented in Fig. 1. The ranking of variables according to all considered measures is the same:

$x_4, x_3, x_2, x_1$ (in descending order). We note that $S_i = Q_i^{(2)}$ at $\alpha = 0.5$ in agreement with Theorem 1.

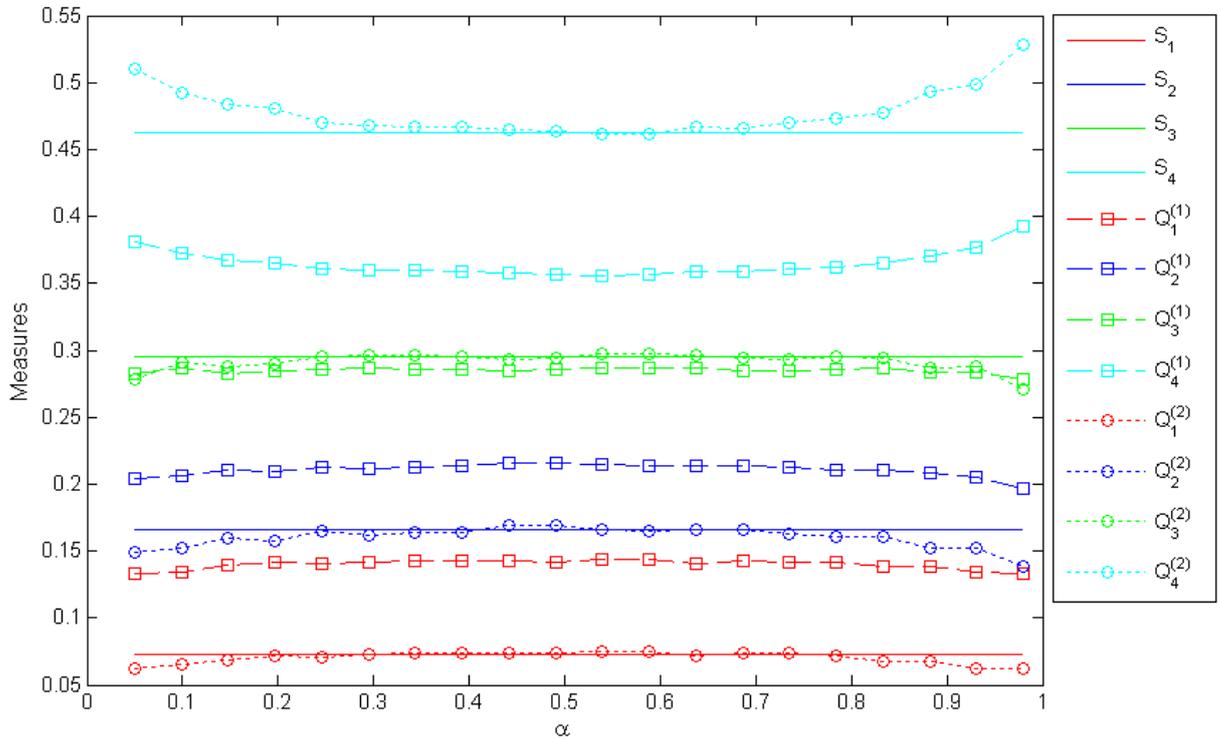

Fig. 1. Values of $Q_i^{(1)}$ and $Q_i^{(2)}$ versus $\alpha$ and values of Sobol' sensitivity indices. Test 1.

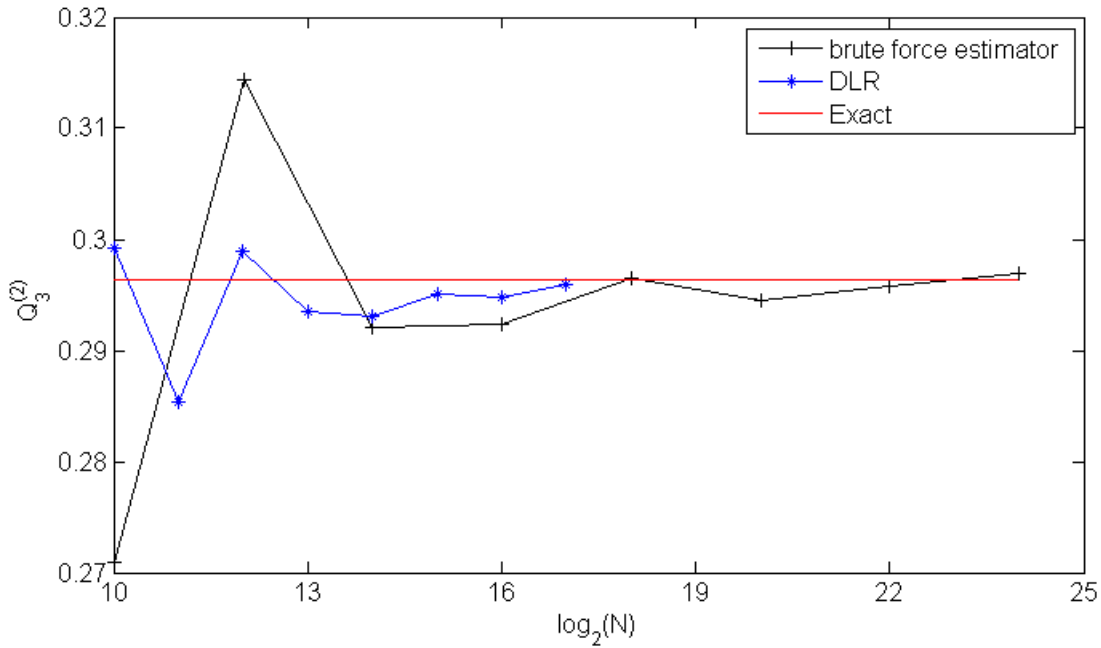

Fig. 2. Convergence of the MC estimators of $Q_3^{(2)}$. Values of $Q_3^{(2)}$ for input 3 at $\alpha = 0.5$ are obtained using the brute force estimator (23) and DLR (26). Test 1.

The convergence comparison of the brute force (23) and DLR (26) MC estimators show that the brute force estimator is much less efficient than DLR (Fig. 2). Further we present only the results obtained with the DLR method.

**Test 2.** Consider the following model $Y = x_1 - x_2 + x_3 - x_4$, where inputs $x_1, x_2, x_3, x_4$ are independent with the following distribution $x_i \sim Exp(\lambda = 1)$. In this case PDF of the output $Y$ is a Laplace distribution $f_Y(y) = \exp(-|2y|)/4$ shown in Fig. 3 (a).

It is possible to find analytical values for Sobol' sensitivity indices $S_i = S_i^{tot} = 0.25$, $i = 1, 2, 3, 4$, and Borgonovo's measure $\delta_i = 0.197$, $i = 1, 2, 3, 4$. Clearly, $x_i$, $i = 1, 2, 3, 4$ have the same influence on $Y$. The values of quantile measures versus $\alpha$ are shown in Fig. 3(b). We notice that dependence of quantile measures versus $\alpha$ is non-linear and non-monotonic. The ranking of variables using $Q_i^{(1)}$ and $Q_i^{(2)}$ measures is different comparing with that of $S_i$ (or $\delta_i$). For small $\alpha$ values $x_2$ has more effect on the quantile of the output than $x_1$, while for large $\alpha$ values the situation is reversed: $x_2$ has less effect on the quantile of the output than $x_1$.

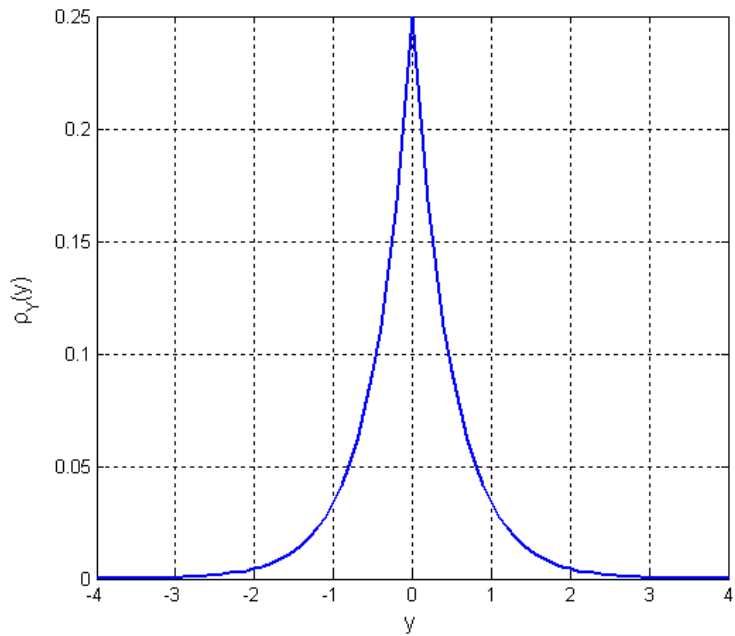

(a)

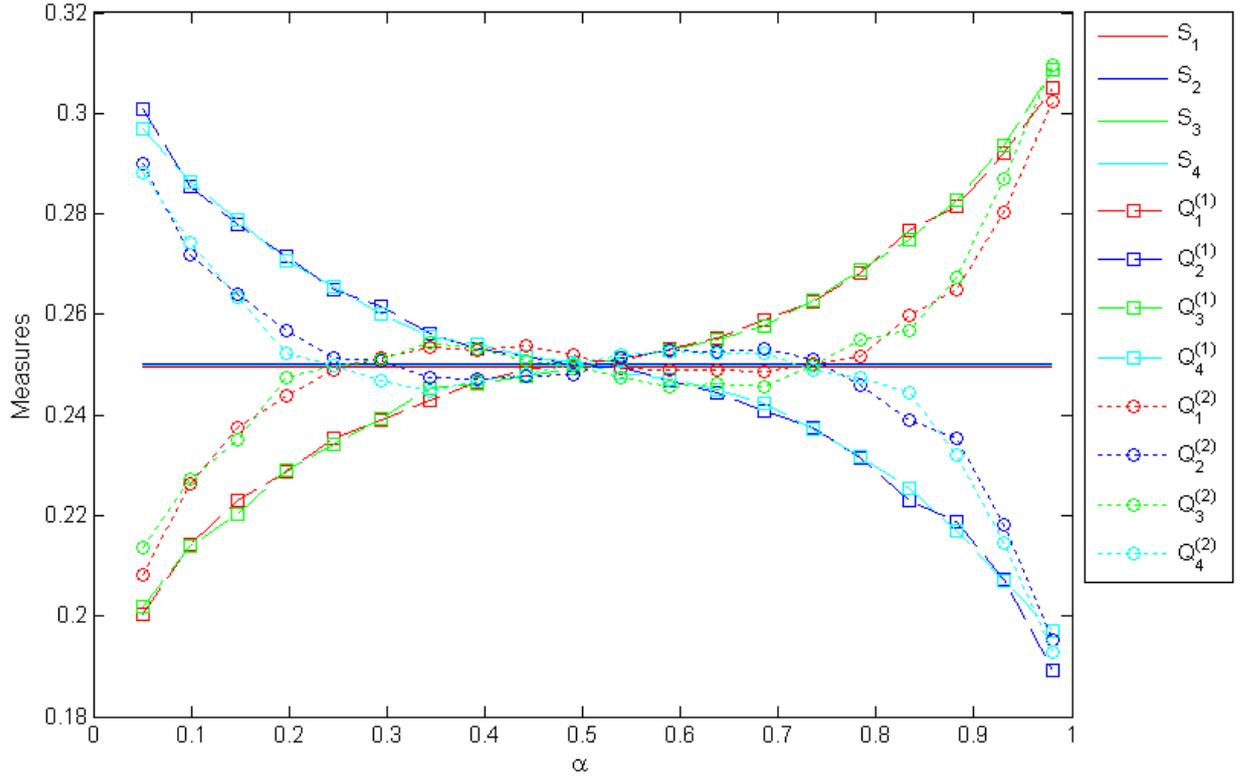

(b)

Fig. 3. The PDF of output Y (a); values of $Q_i^{(1)}$ and $Q_i^{(2)}$ versus $\alpha$ and values of Sobol' sensitivity indices (b). Test 2.

**Test 3.** Consider the following model

$$Y = x_1x_3x_5 + x_1x_3x_6 + x_1x_4x_5 + x_1x_4x_6 + x_2x_3x_4 + x_2x_3x_5 + x_2x_4x_5 + x_2x_5x_6 + x_2x_4x_7 + x_2x_6x_7.$$

All seven variables are independent lognormal with the mean values 2, 3, 0.001, 0.002, 0.004, 0.005 and 0.003 for $x_i$, $i=1,..,7$ respectively. All the standard deviations are equal to 0.4214. This test case was considered in [18, 19]. Values of $S_i$, $S_i^{tot}$ and $\delta_i$ were computed using QMC estimates at $N=2^{15}$. Their values are presented in Table 1.

Table 1. Values of sensitivity measures. Test 3

|  | $x_1$ | $x_2$ | $x_3$ | $x_4$ | $x_5$ | $x_6$ | $x_7$ |
|---|---|---|---|---|---|---|---|
| $S_i$ | 0.0350 | 0.331 | 0.0157 | 0.0858 | 0.174 | 0.221 | 0.0477 |
| $S_i^{tot}$ | 0.0430 | 0.395 | 0.0186 | 0.1000 | 0.215 | 0.265 | 0.0640 |

| $\delta_i$ | 0.0946 | 0.238 | 0.0711 | 0.127 | 0.171 | 0.192 | 0.0993 |

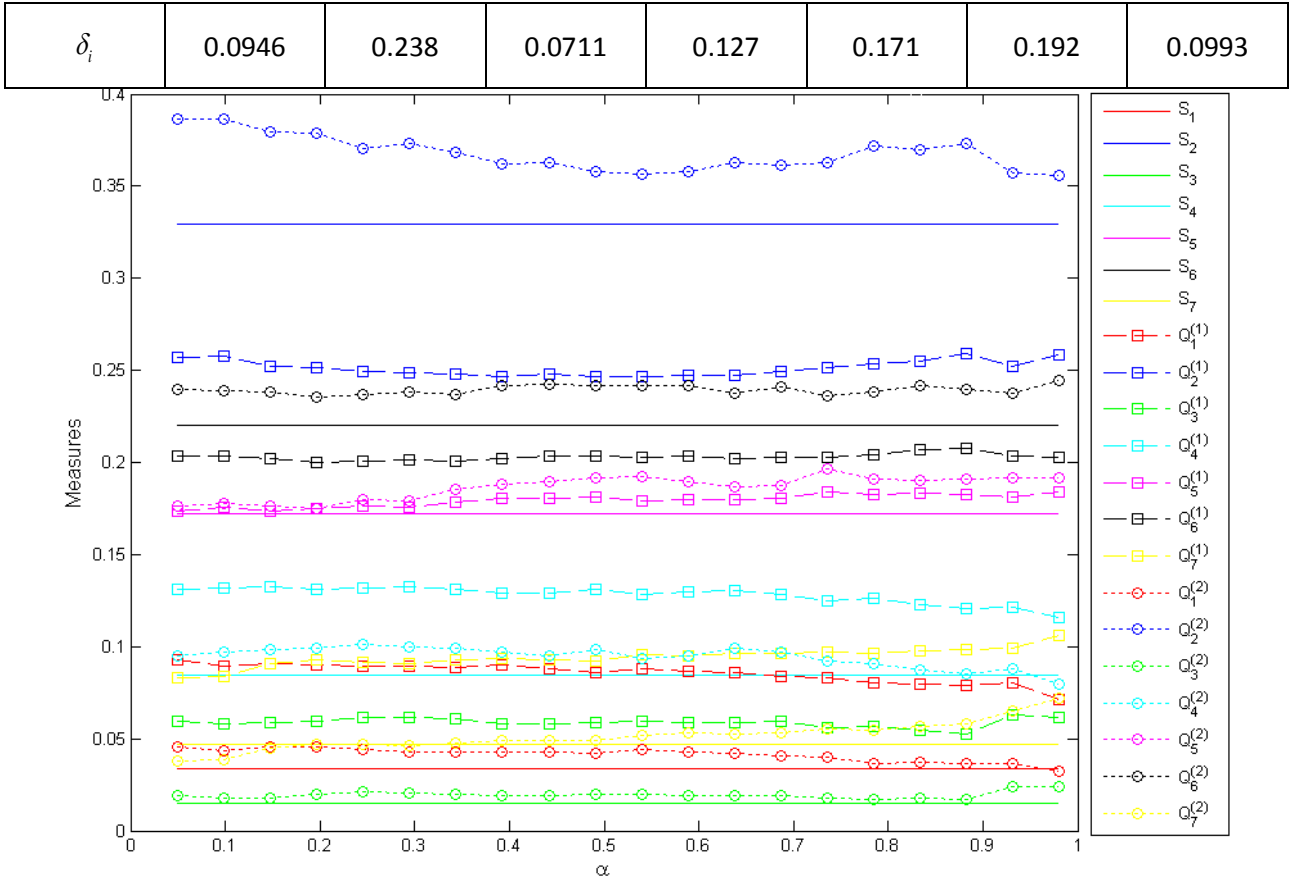

Fig. 4. Values of $Q_i^{(1)}$ and $Q_i^{(2)}$ versus $\alpha$ and values of Sobol' sensitivity indices. Test 3.

The comparison of $Q_i$ and $S_i$ is shown in Fig. 4. The ranking of variables is the same for all considered measures ($S_i$, $S_i^{tot}$ $Q_i$ and $\delta_i$): $x_2, x_6, x_5, x_4, x_7, x_1, x_3$ (in descending order). The dependence of $Q_i(\alpha)$ on $\alpha$ is rather week.

**Test 4**: The Ishigami function $f(\mathbf{x}) = \sin(x_1) + 7(\sin x_2)^2 + 0.1 x_3^4 \sin(x_1)$ is often used in GSA for illustration purposes [1]. Here $x_i$, $i = 1, 2, 3$ are uniformly distributed on the interval $[-\pi, \pi]$. The Sobol's sensitivity indices and Borgonovo's measure have the following values: $S_i$ = {0.314, 0.442, 0.0}, $S_i^{tot}$ = {0.558, 0.442, 0.244}, $\delta_i$ = {0.229, 0.389, 0.0926} for $x_i$, $i = 1, 2, 3$.

This function has a multimodal PDF of the output $Y$ which contains two peaks shown in Fig. 5 (a). It causes highly nonlinear behavior of $Q_i$ versus $\alpha$. Values of $Q_i$ and $S_i$ are shown in Fig. 5 (b). The ranking of variables determined by $S_i$ and $\delta_i$ is the same: $x_2, x_1, x_3$ (in descending order). The ranking of variables determined by $S_i$ and $Q_i$ is the same for $\alpha \in [0.06, 0.96]$. For $\alpha < 0.06$ and $\alpha > 0.96$ the ranking of variables determined by $Q_i$ is different from that defined by $S_i$.

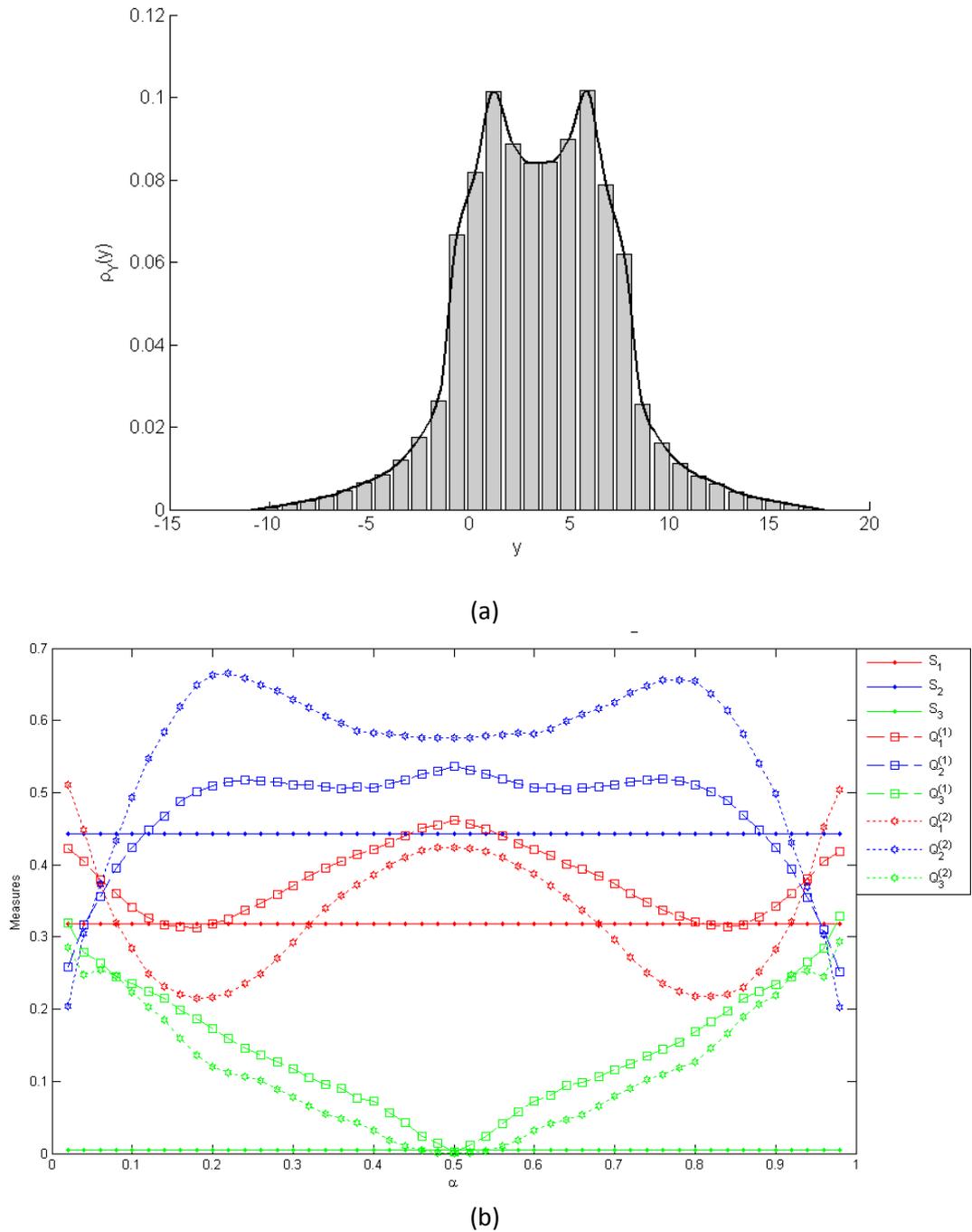

(a)

(b)

Fig. 5. The PDF of output Y (a); values of $Q_i^{(1)}$ and $Q_i^{(2)}$ versus $\alpha$ and values of Sobol' sensitivity indices (b). The Ishigami function.

### 7.2. Correlated inputs

All presented formulas can be used in the case of correlated input variables. Estimation of Sobol' global sensitivity indices $S_i$, $S_i^{tot}$ for models with dependent variables was proposed in [20].

**Test 5.** Consider the following model

$$Y = x_1 x_3 + x_2 x_4, \tag{31}$$

where $(x_1, x_2, x_3, x_4) \sim N(\boldsymbol{\mu}, \Sigma_x)$ with $\boldsymbol{\mu} = (0, 0, \mu_3, \mu_4)$ and $\Sigma_x = \begin{bmatrix} \sigma_1^2 & \sigma_{12} & 0 & 0 \\ \sigma_{12} & \sigma_2^2 & 0 & 0 \\ 0 & 0 & \sigma_3^2 & \sigma_{34} \\ 0 & 0 & \sigma_{34} & \sigma_4^2 \end{bmatrix}$.

This test case was considered in [20], where the analytical values of the first and total order indices were presented:

$$S_1 = \frac{\sigma_1^2 \left(\mu_3 + \mu_4 \rho_{12} \frac{\sigma_2}{\sigma_1}\right)^2}{D}, \quad S_1^{tot} = \frac{\sigma_1^2 (1-\rho_{12}^2)(\sigma_3^2 + \mu_3^2)}{D},$$

$$S_2 = \frac{\sigma_2^2 \left(\mu_4 + \mu_3 \rho_{12} \frac{\sigma_1}{\sigma_2}\right)^2}{D}, \quad S_2^{tot} = \frac{\sigma_2^2 (1-\rho_{12}^2)(\sigma_4^2 + \mu_4^2)}{D},$$

$$S_3 = 0, \quad S_3^{tot} = \frac{\sigma_1^2 \sigma_3^2 (1-\rho_{34}^2)}{D},$$

$$S_4 = 0, \quad S_4^{tot} = \frac{\sigma_2^2 \sigma_4^2 (1-\rho_{34}^2)}{D},$$

where $\rho_{ij} = \frac{\sigma_{ij}}{\sigma_i \sigma_j}$ and $D = \sigma_1^2 (\sigma_3^2 + \mu_3^2) + \sigma_2^2 (\sigma_4^2 + \mu_4^2) + 2\sigma_{12}(\sigma_{34} + \mu_3 \mu_4)$.

For numerical test we used the following parameters: $\boldsymbol{\mu} = (0, 0, 250, 400)$ and

$$\Sigma_x = \begin{bmatrix} 16 & 2.4 & 0 & 0 \\ 2.4 & 4 & 0 & 0 \\ 0 & 0 & 4 \cdot 10^4 & -1.8 \cdot 10^4 \\ 0 & 0 & -1.8 \cdot 10^4 & 9 \cdot 10^4 \end{bmatrix}.$$

The numerical values of Sobol' sensitivity indices and Borgonovo's measure are given in Table 2.

Table 2. Values of sensitivity measures $S_i$, $S_i^{tot}$, $\delta_i$. Test 5

|  | $x_1$ | $x_2$ | $x_3$ | $x_4$ |
|---|---|---|---|---|
| $S_i$ | 0.507 | 0.399 | 0 | 0 |
| $S_i^{tot}$ | 0.492 | 0.300 | 0.192 | 0.108 |
| $\delta_i$ | 0.3379 | 0.3012 | 0.0637 | 0.0422 |

The ranking of variables is the same for all considered measures $S_i$, $S_i^{tot}$ and $\delta_i$ : $x_1, x_2, x_3, x_4$ (in descending order).

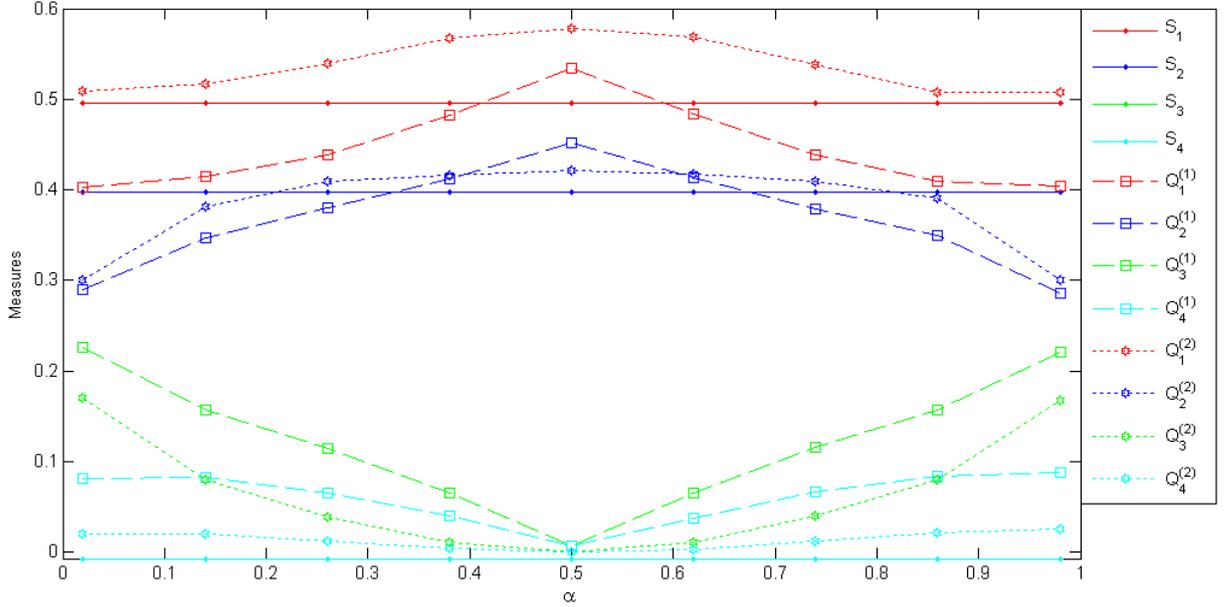

Fig. 6. Values of $Q_i^{(1)}$ and $Q_i^{(2)}$ versus $\alpha$ and values of $S_i$ Test 5. ( $\rho_{34} = -0.3$ )

Comparison of measures $S_i$ and $Q_i$ is presented in Fig. 6. Measures $Q_i$ reach their maximum: *i*=1, 2 and minimum: *i*=3, 4 at $\alpha$ =0.5. It is interesting to note that $Q_i^{(1)}$ and $Q_i^{(2)}$ (*i*=3, 4) grow substantially at very small and very large values of $\alpha$ with $Q_3^{(1)}$ and $Q_3^{(2)}$ almost reaching the values of $Q_2^{(1)}$ and $Q_2^{(2)}$.

## 8. Practical case studies

In this section we consider applications of proposed measures to practical test cases related to structural safety.

**Case study 1. Roof Truss structure.** Consider a roof truss structure shown in Fig. 7 (a). The top boom and the compression bars are reinforced by concrete, while the bottom boom and the tension bars are made of steel.

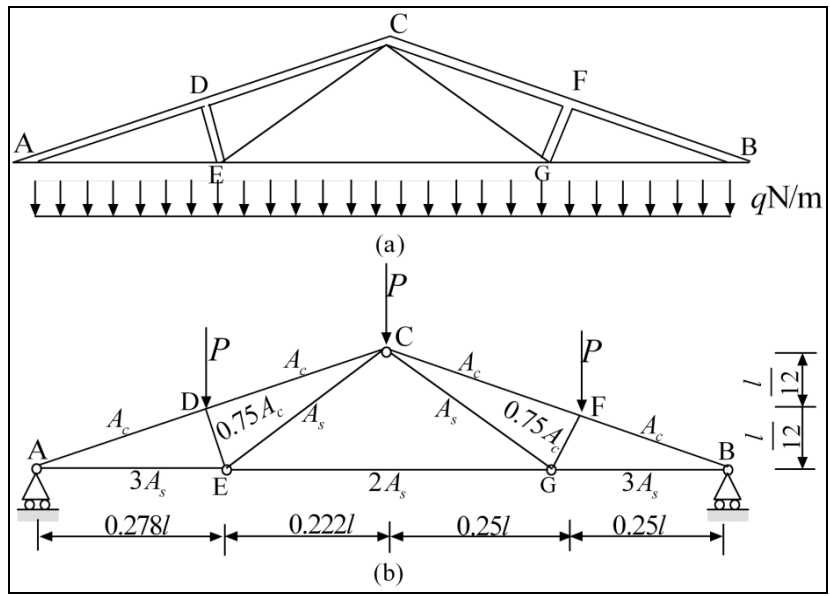

Fig. 7. Roof truss structure (a); distribution of loads and dimensions (b).

Uniformly distributed load *q* can be transformed into the nodal load *P=qL/4*, where *L* is the length of the steel bar (Fig.7 (b)). The perpendicular deflection $\Delta_C$ of node C can be is obtained through basic structural mechanics. It the following function of the input random variables $A_C, A_S, E_C, E_S$ representing respectively sectional areas and elastic moduli of the concrete and steel bars:

$$\Delta_C = \frac{qL^2}{2}\left(\frac{3.81}{A_C E_C} + \frac{1.13}{A_S E_S}\right).$$

Considering as the safety criteria the deflection $\Delta_C$ not exceeding an admissible maximal deflection 3 cm, the performance response model can be constructed by using the limit state function $g(X) = 0.03 - \Delta_C$. We assume that all the input variables are normally distributed with distribution parameters given in Table 4.

Table 3. Roof truss structure. Values of distribution parameters.

| Variable $X$ | Mean $\mu$ | Standard Deviation $\sigma$ |
|---|---|---|
| $q$ (N/m) | 20000 | 140 |
| $L$ (m) | 12 | 0.12 |
| $A_S$ (m²) | 9.82×10⁻⁴ | 5.89×10⁻⁵ |
| $A_C$ (m²) | 0.04 | 0.0048 |
| $E_S$ (N/m²) | 2×10¹¹ | 1.2×10¹⁰ |
| $E_C$ (N/m²) | 3×10¹⁰ | 1.8×10⁹ |

Table 4. Roof truss structure. Values of sensitivity measures $S_i$, $S_i^{tot}$, $\delta_i$, $Q_i^{(1)}$ and $Q_i^{(2)}$. Values of $Q_i^{(1)}$ and $Q_i^{(2)}$ are presented at $\alpha$ close to 0, ranking is given in brackets.

| $x$ | $S_i$ | $S_i^{tot}$ | $\delta_i$ | $Q_i^{(1)}$ | $Q_i^{(2)}$ |
|---|---|---|---|---|---|
| $q$   | 0.453 (1) | 0.463 (1) | 0.271 (1) | 0.317 (1) | 0.499 (1) |
| $L$   | 0.034 (6) | 0.037 (6) | 0.068 (6) | 0.083 (6) | 0.034 (6) |
| $A_S$ | 0.141 (3) | 0.144 (3) | 0.131 (3) | 0.159 (4) | 0.124 (4) |
| $A_C$ | 0.189 (2) | 0.189 (2) | 0.150 (2) | 0.189 (2) | 0.174 (2) |
| $E_S$ | 0.141 (4) | 0.141 (3) | 0.131 (3) | 0.162 (3) | 0.128 (3) |
| $E_C$ | 0.039 (5) | 0.044 (5) | 0.072 (5) | 0.091 (5) | 0.041 (5) |

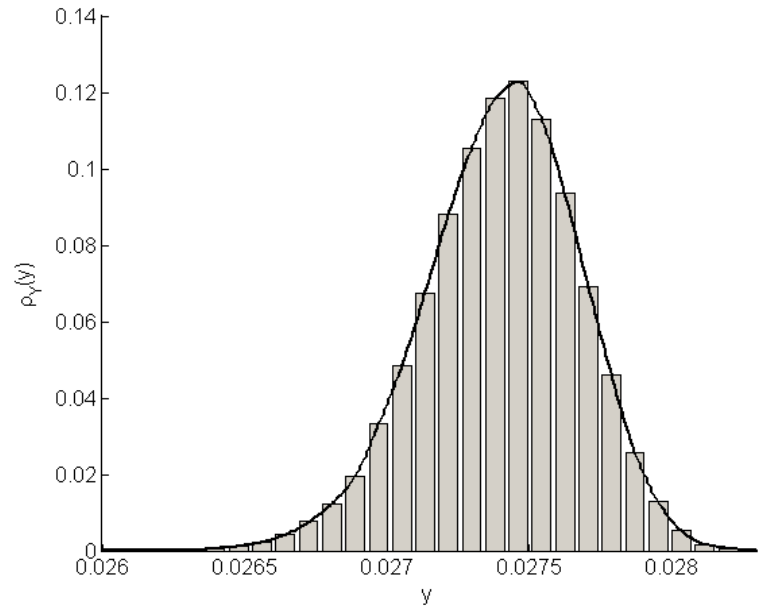

(a)

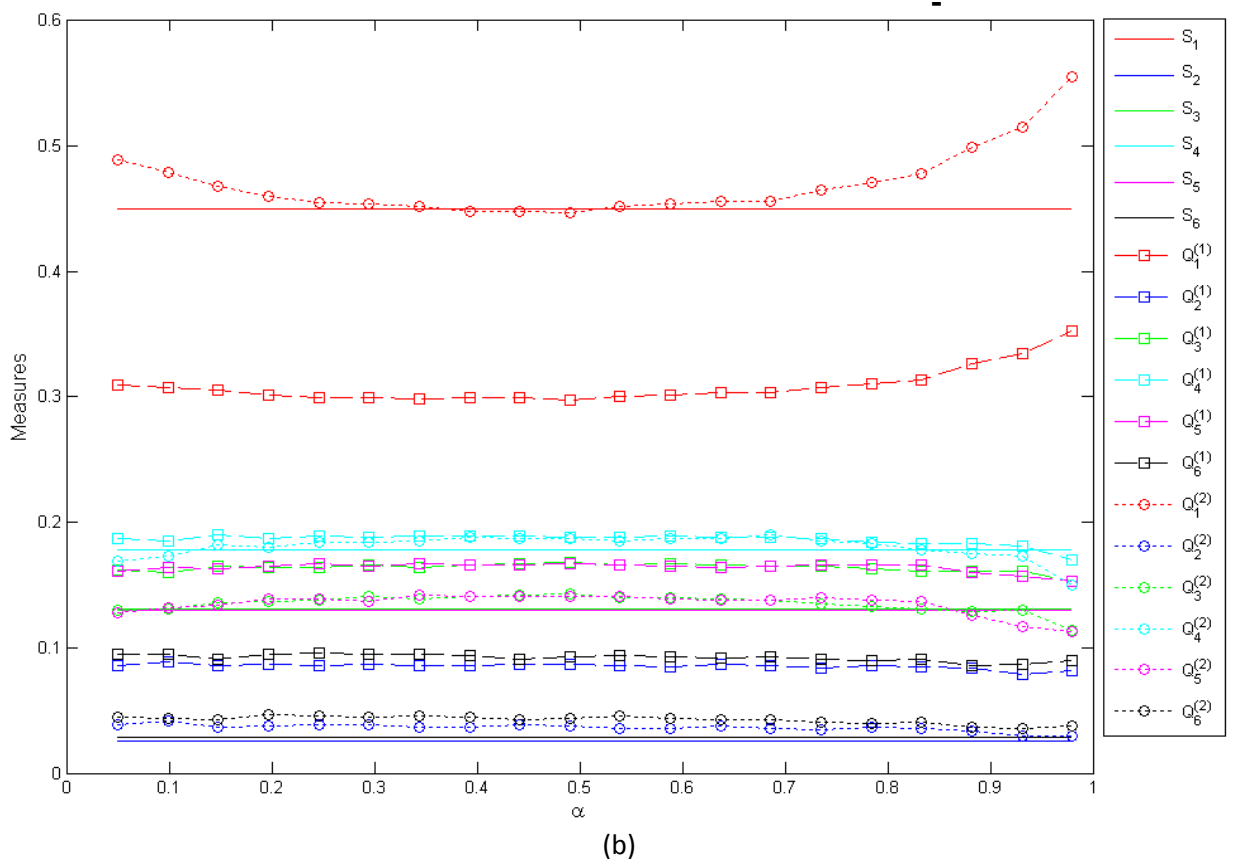

(b)

Fig. 8. The PDF of output $g(X)$ (a); values of $Q_i^{(1)}$ and $Q_i^{(2)}$ versus $\alpha$ and values of Sobol' sensitivity indices (b). Roof truss structure.

PDF of the model random response $g(X)$ is skewed (Fig. 8 (a)) which causes non monotonic nonlinear behavior of $Q_i$ versus $\alpha$ (Fig. 8 (b)). The ranking of variables determined by $S_i$, and $\delta_i$ is the same with $q$ being the most important and $L$ being the least significant input, respectively. The ranking of variables using $Q_i^{(1)}$ and $Q_i^{(2)}$ measures is different comparing with that of $S_i$ (or $\delta_i$) for only for inputs 3 ($A_s$) and 4 ($E_s$), although the values of sensitivity indices of these two inputs differ in less than 3%.

**Case study 2. Creep-fatigue failure model.** Creep is one of the principal damage mechanisms for materials operating at elevated temperatures. The design of a structure or component is reliable when the total creep–fatigue damage is less than the allowable damage. Creep–fatigue failure can be defined by the rule $D_c + D_f > D_{cr}$, where $D_c$ and $D_f$ correspond to the creep damage and fatigue damage respectively, $D_{cr}$ is the critical damage that is determined according the experimental test results. It is dependent on the material.

Reliability analysis of creep and fatigue failure can be assessed by computing the probability of failure of systems with respect to some criterion function $g(X)$. Further we consider a probabilistic model for reliability analysis of creep and fatigue of materials proposed in [21], where the following nonlinear creep-fatigue failure criterion function based on experimental data using linear damage accumulation rule was proposed:

$$g(N_c, N_f, n_c, n_f, \theta_1, \theta_2) = D_{cr} - (D_c + D_f) = 2 - e^{\theta_1 D_c} + \frac{e^{\theta_1} - 2}{e^{-\theta_2} - 1}(e^{-\theta_2 D_c} - 1) - D_f.$$

Here $D_c = n_c/N_c$, $D_f = n_f/N_f$, $\theta_1$ and $\theta_2$ are the parameters obtained from the experimental results, $N_c$ and $N_f$ correspond to the creep life and fatigue life respectively, $n_c$ and $n_f$ are the number of the creep and fatigue loads cycles. It is assumed that the creep and fatigue life times of material follow log-normal distributions while parameters $\theta_1$, $\theta_2$ follow normal distributions. Values of the distribution parameters are given in Table 5.

Table 5. Creep-fatigue failure model. Values of distribution parameters.

| Variable $X$ | Mean $\mu$ | Coefficient of variation $C = \sigma/\mu$ | Distribution type |
|---|---|---|---|
| $N_c$ | 5490 | 0.20 | Log-normal |
| $N_f$ | 17100 | 0.20 | Log-normal |
| $n_c$ | 549 | 0.20 | Log-normal |
| $n_f$ | 4000 | 0.20 | Log-normal |
| $\theta_1$ | 0.42 | 0.20 | Normal |
| $\theta_2$ | 6.0 | 0.20 | Normal |

Table 6. Creep-fatigue failure model. Values of sensitivity measures $S_i$, $S_i^{tot}$, $\delta_i$, $Q_i^{(1)}$ and $Q_i^{(2)}$. Values of $Q_i^{(1)}$ and $Q_i^{(2)}$ are presented at $\alpha = 0.5$, ranking is given in brackets.

| $x$ | $S_i$ | $S_i^{tot}$ | $\delta_i$ | $Q_i^{(1)}$ | $Q_i^{(2)}$ |
|---|---|---|---|---|---|
| $N_c$ | 0.284 (1) | 0.288 (1) | 0.197 (1) | 0.242 (1) | 0.301 (1) |
| $N_f$ | 0.0792 (3) | 0.0792 (3) | 0.091 (3) | 0.110 (3) | 0.062 (3) |
| $n_c$ | 0.284 (1) | 0.288 (1) | 0.197 (1) | 0.242 (1) | 0.301 (1) |
| $n_f$ | 0.0792 (3) | 0.0792 (3) | 0.091 (3) | 0.110 (3) | 0.062 (3) |
| $\theta_1$ | 0.0378 (4) | 0.040 (4) | 0.068 (4) | 0.096 (4) | 0.049 (4) |
| $\theta_2$ | 0.232 (2) | 0.236 (2) | 0.176 (2) | 0.201 (2) | 0.225 (2) |

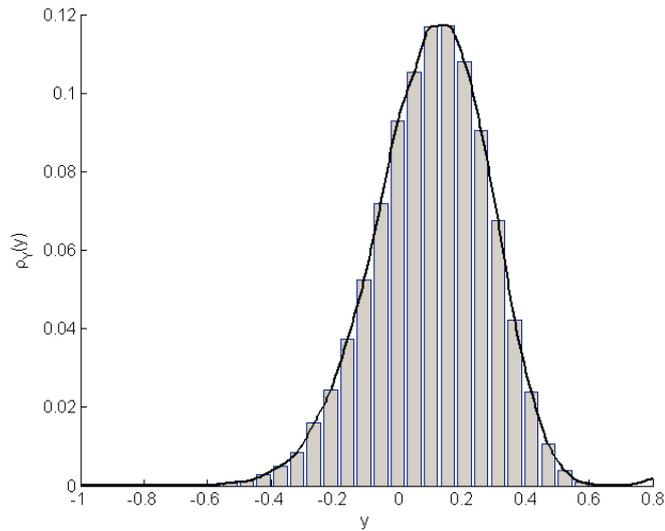

(a)

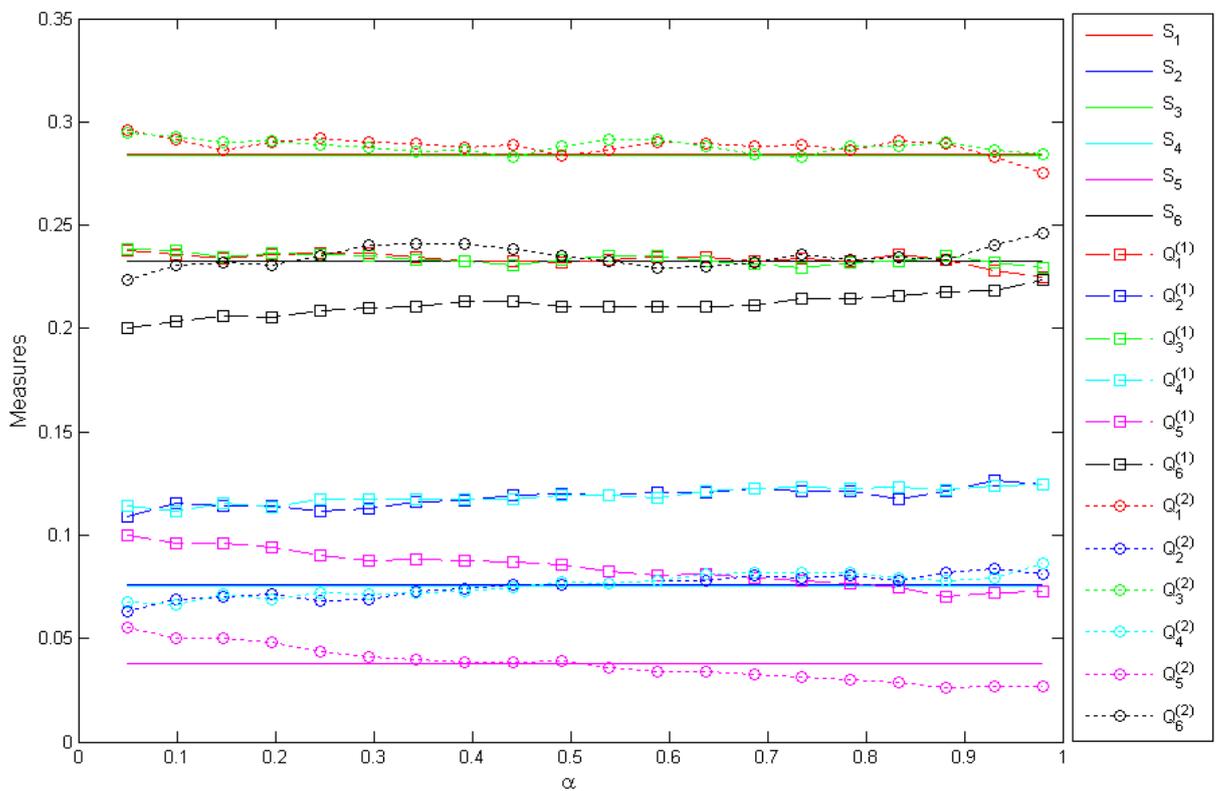

(b)

Fig. 9. The PDF of output $g(X)$ (a); values of $Q_i^{(1)}$ and $Q_i^{(2)}$ versus $\alpha$ and values of Sobol' sensitivity indices (b). Creep-fatigue failure model

PDF of the output $g(X)$ is symmetric (Fig. 9 (a)). It results in practically linear behavior of $Q_i$ versus $\alpha$ (Fig. 9 (b)). The ranking of variables determined by all sensitivity measures is the same with $N_c$ being the most important and $\theta_1$ being the least significant input, respectively.

## 9. Conclusions

We proposed novel sensitivity measures based on quantiles of model outputs for global sensitivity analysis of problems in which $\alpha$-th quantiles are the functions of interest. Such measures can also be useful for problems in which the analysts are interested in ranking of inputs contributing to the extreme values of the output. In the case of linear additive models with normally distributed inputs quantile based measures have a direct link with Sobol' main effect sensitivity indices. We considered the double loop reordering and the brute force MC estimators. It was shown that the double loop reordering estimator has much better convergence properties than the brute force estimator.

The efficiency and accuracy of the proposed measures were illustrated in application examples. Different models with both independent and dependent (correlated) inputs were considered. It was shown that depending on the model and the type of inputs PDF's quantile based measures can show a non-linear and non-monotonic behavior versus $\alpha$-th quantile. These measures can be especially useful in the case of highly skewed or fat-tailed output distributions.

## Acknowledgement

The authors would like to thank E. Borgonovo for his constructive comments and N. Shah for his support and interest in this work. The financial support by the EPSRC grant EP/H03126X/1 is gratefully acknowledged.